\documentclass[12pt,a4paper,final]{iopart}
\expandafter\let\csname equation*\endcsname=\relax
\expandafter\let\csname endequation*\endcsname=\relax
\usepackage{amsmath}
\usepackage{amssymb}
\usepackage{float}
\usepackage{color}
\usepackage{graphicx}
\usepackage{dcolumn}
\usepackage{bm}
\usepackage{xcolor}
\usepackage{multirow}
\usepackage{threeparttable}
\usepackage{adjustbox}
\usepackage{longtable}
\usepackage{comment}
\usepackage{array}
\usepackage[normalem]{ulem}
\usepackage[breaklinks=true,colorlinks=true,linkcolor=red,urlcolor=blue,citecolor=blue]{hyperref}
\begin{document}



\title[Phys. Scr. (2022)]{Systematic study of the effect of individual rotational energy levels on the fusion cross-section of \texorpdfstring{$^{16}O$}--based reactions of range $480 \le {Z_PZ_T} \le 592$}

\author{Nishu Jain$^{1}$, M. Bhuyan$^{2}$ and Raj Kumar$^{1}$}
\address{$^1$ School of Physics and Materials Science, Thapar Institute of Engineering and Technology, Patiala-147004, Punjab, India}
\address{$^2$ Center for Theoretical and Computational Physics, Department of Physics, Faculty of Science, University of Malaya, Kuala Lumpur 50603, Malaysia}
\ead{\href{nishujain1003@gmail.com}{nishujain1003@gmail.com}} 
\ead{\href{bunuphy@um.edu.my}{bunuphy@um.edu.my}}	
\ead{\href{rajkumar@thapar.edu}{rajkumar@thapar.edu}}  
\vspace{10pt}
\begin{indented}
	\item[]September 2022
\end{indented}

\begin{abstract}
\noindent
In heavy-ion fusion reactions, the enhancement in the sub-barrier fusion cross-section has been observed as compared to the 1-Dimensional barrier penetration model due to the coupling of many degrees of freedom to the relative motion. This enhancement can be explained theoretically by including nuclear structure effects like deformation and the coupling of relative motion among two colliding nuclei. The present work aims to investigate the effect of individual rotational energy levels on the fusion cross-sections for $^{16}$O-based reaction systems, namely, $^{16}$O + $^{182,184,186}$W,  $^{16}$O + $^{176,180}${Hf}, $^{16}$O + $^{174,176}${Yb}, $^{16}$O + $^{166}${Er}, $^{16}$O + $^{148,152,154}$Sm, $^{16}$O + $^{150}$Nd at energies below the fusion barrier. Using the CCFULL code, the effect of low-lying rotational energy levels on the fusion cross-section for $^{16}$O induced reactions has been investigated at energies below and around the Coulomb barrier. The calculations are performed by assuming the fixed value of diffuseness parameter $a_{0}=0.65$ fm in the Woods-Saxon nuclear potential and the other two parameters are optimised by fitting the experimental data at the above barrier. Here we have determined the $V_0$ and $r_0$ as a function of $Z_PZ_T$, where experimental cross-sections are available. From our calculations, it is observed that the hexadecapole deformation ($\beta_4$) with different magnitudes has a significant influence on the fusion cross sections. For the case of the $+ve$ value of $\beta_4$, beyond $10^+$, the rotational levels cease to contribute significantly and also there is a significant difference between the contribution of sequential channels. On the other hand, in the case of -ve $\beta_4$, up to $6^+$ levels contribute significantly.  Furthermore, we have established an algebraic systematic of fitting, which one can use to determine the parameters $V_0$, $r_0$ of Woods-Saxon nuclear potential within the range of $Z_PZ_T$ lie in between $480 \le {Z_PZ_T} \le 592$.
\end{abstract}

\section{Introduction}
Over the last few decades, various theoretical and experimental efforts have been centred on exploring the role of nuclear structure in the reaction dynamics \cite{dasg98,beck88,stea86}. During the fusion process, the collision of an incident projectile and a target may lead to the formation of a compound nucleus through a quantum tunnelling process across the fusion barrier \cite{dasg98,bala98,cant06,back14}. Heavy-ion fusion reactions are accomplished in the crucial role of extending the nuclear chart and synthesizing the heavy elements. Furthermore, some heavy-ion fusion processes in lighter mass systems are pivotal to the reaction channels that govern the elemental synthesis in stellar environments and energy production \cite{beck88,stea86,brom84}. Experimentally, it has been observed that there is a large enhancement in the sub-barrier fusion cross-sections by considering the one-dimensional barrier penetration model \cite{leig95,bier96,stok81,stef95,newt04,zagr03} using bare nucleon-nucleon potentials. Such sub-barrier fusion enhancement could be elucidated by the coupling of relative motion degrees of freedom with the internal degrees of freedom such as static deformations (rotational nuclei) \cite{bier96,stok81}, vibrational effect in the nuclear surface (spherical nuclei) \cite{Newt01,stef95a,stef84}, and the neck formation \cite{krap83}. However, the effects of nucleon transfer on fusion below the barrier have not yet been completely separated \cite{back14,deni00,lagy97,trip01,vija22,stef00,Zagr07,stef07,suni10,khus19,deb22}. Several works \cite{dasg98,Newt01,rowl91,naji19,kaur16,kaur18,rumi01,Mort94,hagi12} have discussed the coupling between the rotational state of the target nuclei. The internal degrees of freedom coupled with the relative motion of colliding nuclei lowers the barrier. The significant decrease in the fusion cross-section with respect to the experimental data, at energies below the Coulomb barrier, is also known as fusion hindrance \cite{bala98}. A certain amount of flux allows the projectile to tunnel through the fusion barrier and fuse well into the target \cite{dasg98,leig95} mainly at below barrier energies. Therefore, coupled-channel calculations become a benchmark theoretical tool to understand fusion reaction dynamics.

Nuclear fusion is a complicated phenomenon because of the involvement of many nucleons. Therefore, it is tedious to handle the interaction potential between two colliding nuclei, which play a key role in describing the fusion reaction dynamics. In general, the interaction potential comprises long-range repulsive Coulomb potential, centrifugal interaction, and short-range attractive nuclear potential terms. Unlike Coulomb and centrifugal potentials, the nuclear potential is not well established till now. In the past few decades, various efforts have been made to provide a simple and accurate form of the nuclear interaction potential \cite{satc79,khoa00,nege82,umar16,roye00,roye02,bloc77}. Several fits of the nuclear potentials exist in the literature \cite{zamr06,mohr19,chen19,cina21,deni22} which either kept the coupling off and/or included in the Woods-Saxon form while using the other potentials. The traditional Woods-Saxon potential has been prominent and widely used to probe the heavy-ion nuclear fusion dynamics \cite{rowl91,gaut15a,Saga07,Esbe10,Stef09,gaut16,gaut14,ibra13}. The Woods-Saxon potential consists of three parameters namely; potential depth, range and diffuseness parameter. The diffuseness parameter is an essential component in the parameterization of Woods-Saxon nuclear potential, as it describes the slopes of the potential in the tail region, where fusion begins \cite{Gaut15}. Remarkably, the small value of the diffuseness parameter $a_{0} = 0.65 fm$ is most suitable for a good description of experimental data \cite{newt04,hagi04} in the elastic scattering analysis while the large values of the diffuseness parameter ranging from $a =$ 0.75 to 1.5 $fm$ provide the best fit for several excitation functions in the above barrier region to explore the experimental data \cite{bala98,cant06,newt04,hagi04,mukh07}. As a result, the Woods-Saxon nuclear potential parameters are employed to study the elastic scattering and heavy-ion fusion processes in the Coupled channel approach (CCFULL), which provides a good description of below barrier fusion \cite{hagi04,mukh07}.

It is known that the nuclear reactions are adequately affected by the entrance channel parameters and the internal structure such as mass asymmetry, deformation, and orientation of the colliding nuclei \cite{shil08,pras10,fer1991,Rhoa83}, which are remarkably governed to influence the anticipation of a compound nucleus system. The significance of shape degrees of freedom in sub-barrier fusion enhancement has been investigated experimentally using the fusion of two colliding nuclei \cite{Gaut19}. As we know, CCFULL is one of the computational codes used to explore fusion dynamics, where a full description of the Woods-Saxon nuclear potential parameters ($V_0$, $r_0$ and $a_0$) for colliding nuclei is essential. To understand the nuclear interaction potential for a recently synthesized or anticipated target nuclei for $^{16}$O- induced reactions, one must fit the main ingredient parameters for the nuclear potential as per the shape degrees of freedom. We will also introduce an algebraic fitting of $V_0$ and $r_0$ by using the available known data. This will be crucial for the theoretical calculations in predicting the fusion characteristics of compound nuclei, which is an essential input for the upcoming experiments.

Furthermore, the study will include the significant role of each rotational energy level (i.e. 2$^+$, 4$^+$, 6$^+$, 8$^+$, 10$^+$ and 12$^+$) in the enhancement of the fusion cross-section mainly at energies below the Coulomb barrier for spherical and deformed nuclei. Here, the fusion cross-section of 12 different $^{16}$O- induced reactions with rotational target nuclei i.e., $^{16}$O+$^{182,184,186}$W, $^{176,180}${Hf}, $^{174,176}${Yb}, $^{166}${Er}, $^{148,152,154}${Sm}, $^{150}${Nd} will be analyzed at sub-barrier energies within the static Woods-Saxon potential with standard diffuseness parameter $a_0$ = 0.65 {\it fm}. We have considered $^{16}$O as spherical \cite{Gaut19,rajb16,rajb14} to investigate the effect of the individual rotational energy level of target nuclei on the fusion cross-section. It is worth mentioning that the low-lying rotational energy levels for the chosen target nuclei follow the sequence of state with $I = 0^+, 2^+, 4^+, 6^+, etc$ and the excitation energies of $I^+$ state is proportional to $I(I+1)^*E_2T/6 $ \cite{kran86}. Moreover, a better channel selection has been made for different signs of $\beta$-values \cite{Mort94,fer1991,Rhoa83,Lemm93,Fern89} for individual energy levels. In addition, a comparison will be made between the resulting theoretical results and the available experimental data for the nuclear reactions under consideration. \\
This paper is organized as follows: A brief description of the theoretical formalism used in this work is given in Section \ref{theory}. The results of the coupled channel calculations are given in detail in Section \ref{results}. Section \ref{result} summarizes and concludes this work.

\section{Theoretical Formalism} 
\label{theory} \noindent
This section provides a brief description of Coupled Channel approach (CCFULL) used in the present study, which provides a reasonable understanding of the nuclear fusion dynamics at energies around the barrier. Within this approach, multidimensional barrier penetration is considered instead of single barrier penetration. This method is used under the effect of coupling of the relative motion with the intrinsic degrees of freedom of the interacting nuclei \cite{bala98,hagi12,hagi99}, mainly for the calculation of mean angular momenta and the fusion cross-sections of the compound nucleus.  The traditional method for addressing the effects of the coupling between relative motion and intrinsic degrees of freedom on fusion is to numerically solve the coupled-channels equations, which includes all relevant channels \cite{Esbe87,rumi99}. The CC equations solve numerically within the Coupled channel approach are given as, 
\begin{eqnarray}
\Bigg[\frac{-\hbar^2}{2\mu} \frac{d^2 }{dr^2} &+& \frac{J(J+1)\hbar^2}{2\mu r^2}+\frac{Z_P Z_T e^2}{r}+V_{N}+\epsilon_n -E_{c.m.}\Bigg] \nonumber \\
&& \times \psi_n (r) + \sum_m V_{nm} (r)\psi_m (r) =0. 
\label{cc1}   
\end{eqnarray}
Here $r$ defines the radial part of relative motion between the participating nuclei and $\mu$ is known as the reduced mass of the colliding system. $\epsilon_n$ is the excitation energy for the $n^{th}$ channel and $E_{c.m.}$ is the bombarding energy in the centre of the mass frame. $V_{N}$ represents the nuclear potential and $V_{nm}$ symbolizes the matrix elements of the coupled Hamiltonian. Since there are several Coupled channel equations, their dimension is also large. Thus, the rotating frame approximation or no-Coriolis approximation is employed to reduce the dimension of Coupled channel equations \cite{hagi12,hagi99,hagi98,muha08}. The CC equations with non-linear coupling are significant in studying the heavy-ion fusion reactions mainly at sub-barrier energies. All these sets of non-linear coupling are taken into account. 

The incoming wave boundary conditions (IWBC) \cite{land84} are also essential for the solution of Coupled channel equation because IWBC or ingoing wave conditions are quite sensitive for the potential pocket of the interaction fusion barrier. The incoming wave of the entrance channel is present inside the barrier at the minimum position (r = $r_{min}$) and the outgoing wave of other channels are present at an infinite position. By including the effect of the dominant intrinsic channels, the fusion cross-sections are calculated as given below:
\begin{eqnarray}
\sigma_J (E)=\sigma_{fus} (E)=\frac{\pi}{k_{0}^{2}}\sum_{J}(2J+1)P_J (E).
\label{fusion}
\end{eqnarray}
Here, the total angular momentum {\it `J'} is substituted in place of `$\ell$' for each channel by applying iso-centrifugal approximation by using the following equation:
\begin{eqnarray}
\langle{\ell}\rangle&=&\sum_{J} J \sigma_J (E) / \sum_{J} \sigma_J (E)
=\Bigg(\frac{\pi}{k_{0}^{2}}\sum_{J} J(2J+1)P_J (E)\Bigg) \nonumber \\
&&   \bigg / \Bigg(\frac{\pi}{k_{0}^{2}}\sum_{J} (2J+1)P_J (E)\Bigg),   
\label{am}
\end{eqnarray}
where $P_J (E)$ is the total transmission coefficient. The Woods-Saxon form of nuclear potential is used to analyze the nuclear structure effects \cite{hagi12,hagi99,hagi98,muha08} and it is defined as
\begin{equation}
    V_{N}= \frac{-V_0}{1+\exp\Big[\big(r_0-R_0\big)/a_0\Big]},  \\ 
    \label{ws}
\end{equation}
where $V_0$, $r_0$, and $a_0$ are the nuclear potential parameter.

In the Coupled channel approach, the rotational coupling with a pure rotor is taken into consideration. One can generate the nuclear coupling Hamiltonian by changing the target radius in the nuclear potential to a dynamical operator, 
\begin{equation}
   R_0 \rightarrow R_0 + \hat{O}= R_0 + \beta_2R_TY_{20}+\beta_4R_TY_{40}.
\end{equation}
Here $R_T$ is $r_{coup}A^{1/3}$ and $\beta_2$ and $\beta_4$ are the quadrupole and hexadecapole deformation parameters of the deformed target nucleus, respectively. Thus, the nuclear coupling Hamiltonian is given by
\begin{equation}
    V_{N}(r,\hat{O})= \frac{-V_0}{1+\exp\Big[\big(r_0-R_0-\hat{O}\big)/a_0\Big]},  \\ 
    \label{ws1}
\end{equation}
To connect the $|n\rangle=|I0\rangle$ and $|m\rangle=|I'0\rangle$ states of the target's ground rotational band, we need matrix elements of the coupling Hamiltonian. These are readily accessible using matrix algebra \cite{Kerm93}. In this algebra, the eigenvalues and eigenvectors of the operator $\hat{O}$, which satisfies
\begin{equation}
    \hat{O}\mid \alpha > = \lambda_\alpha\mid \alpha > 
\end{equation}
This is implemented in the CCFULL program by diagonalizing the matrix $\hat{O}$, whose elements are given by 
\begin{eqnarray}
\begin{aligned}
    \hat{O}_{II'}= \sqrt{\frac{5(2I+1)(2I'+1)}{4\pi}}\beta_2R_T
\begin{pmatrix}
I & 2 & I'\\
0 & 0 & 0 
\end{pmatrix}^2 \\
+ \sqrt{\frac{9(2I+1)(2I'+1)}{4\pi}}\beta_4R_T
\begin{pmatrix}
I & 4 & I'\\
0 & 0 & 0 
\end{pmatrix}^2.  
\end{aligned}
\end{eqnarray}
The nuclear coupling matrix elements are then evaluated as
\begin{eqnarray}
V_{nm}^{(N)}&=& <I0\mid V_N(r,\hat{O})\mid I^{'}0> - V_{N}^{(0)}(r)\delta_{n,m}, \nonumber \\
&&  = \sum_\alpha <I0\mid \alpha><\alpha\mid I^{'}0> V_N(r,\lambda_\alpha)-V_{N}^{(0)}(r)\delta_{n,m}.   
\label{am1}
\end{eqnarray}
The last term is included in the equation to avoid the diagonal component from being counted twice. The linear rotational coupling approximation is used to calculate the Coulomb matrix and explained as \\
\begin{eqnarray}
    \begin{aligned}
        V^C_{R{(I,I')}}=\frac{3Z_PZ_TR^2_T}{5r^3}\sqrt{\frac{5(2I+1)(2I'+1)}{4\pi}}\times \\
        \Bigg( \beta_{2}+\frac{2}{7}\beta^2_2\sqrt{\frac{5}{\pi}}\Bigg)\begin{pmatrix}
I & 2 & I'\\
0 & 0 & 0 
\end{pmatrix}^2 \\+\frac{3Z_PZ_TR^4_T}{9r^5}\sqrt{\frac{9(2I+1)(2I'+1)}{4\pi}}\times\\
\Bigg( \beta_{4}+\frac{9}{7}\beta^2_2\Bigg)\begin{pmatrix}
I & 4 & I'\\
0 & 0 & 0 
\end{pmatrix},
    \end{aligned}
\end{eqnarray}
for rotational coupling. These coupled channel equations are used to calculate the fusion cross-section of the compound nucleus by considering the coupling of all orders as discussed in Sec. \ref{results}.

\begin{table} 
\caption{\label{table1} The parameters  of  Woods-Saxon potential ($V_0$ $\&$ $r_0$), the deformation parameters ($\beta_2 > 0, \beta_4 < 0$) and the excitation energy corresponding to quadrupole deformation of the nuclei \cite{rama01,moll16} used in the coupled channel calculations. }
\centering
{\begin{tabular}{ccc|ccc}
\hline \hline 
 System & $V_0 (MeV)$ & $r_0 (fm)$ & \multicolumn{3}{c}{Target} \\
        &         & & $E_2^+ (MeV)$ & $\beta_2$ & $\beta_4$ \\
\hline
 $^{16}$O+$^{182}$W& 63.899& 1.165& 0.100 &0.2500&-0.066 \\
 $^{16}$O+$^{184}$W& 63.987& 1.178& 0.111& 0.236& -0.093\\
  $^{16}$O+$^{186}$W& 70.0& 1.18& 0.122& 0.221& -0.095\\
  $^{16}$O+$^{176}${Hf}&63.627&1.18&  0.088&0.295& -0.057 \\
  $^{16}$O+$^{180}${Hf}&70.5&1.17& 0.093&0.274& -0.068 \\
  $^{16}$O+$^{174}${Yb}&63.53&1.18& 0.076&0.325& -0.042 \\
  $^{16}$O+$^{176}${Yb}&60.0&1.165& 0.082&0.304& -0.068\\
   \hline
\end{tabular}}
\end{table}
\begin{figure}
\begin{center}
\includegraphics[width=70mm,height=70mm,scale=1.5]{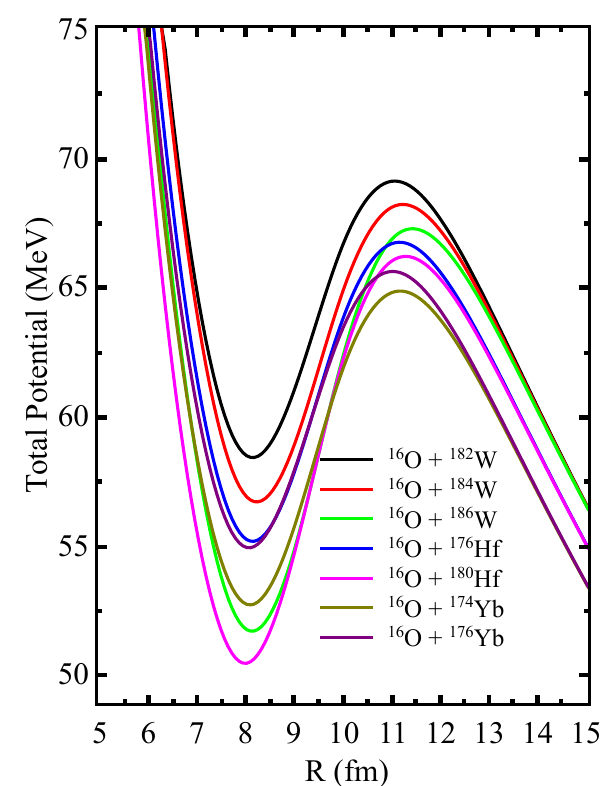}
\vspace{-0.3cm}
\caption{\label{fig. 1} (Color online) The variation of total interaction potential in the respect of separation distance r (fm) for $^{16}$O+$^{182,184,186}$W, $^{16}$O+$^{176,180}$Hf, and $^{16}$O+$^{174,176}$Yb reactions under $\beta_2 > 0$, $\beta_4 < 0$ condition.}
\end{center}
\end{figure}

\section{Results and Discussions}
\label{results} \noindent
The fusion cross-sections have been calculated for $^{16}$O-induced nuclear reactions with different rotational target nuclei, namely, $^{16}$O + $^{182,184,186}$W, $^{16}$O + $^{176,180}${Hf}, $^{16}$O + $^{174,176}${Yb}, $^{16}$O + $^{166}${Er}, $^{16}$O + $^{148,152,154}${Sm}, $^{16}$O + $^{150}${Nd} with Coupled channel calculations by adopting CCFULL code. Various theoretical approaches have been developed to generate the attractive nuclear potential for the description of fusion data over wide energy ranges \cite{Nobr07,Siwe01,Bala97}. In this work, Woods-Saxon parameterization has been taken into account to compute the interaction potential. The standard value of diffuseness parameter $a_0$ = 0.65 fm is used for the considered nuclear reactions \cite{newt04,hagi04}. The deformation parameter $\beta_\lambda$ connected with the transition of multipolarity $\lambda$ were determined from experimental transition probabilities B(E2) by using the following relation:
\begin{equation}
\beta_\lambda=\frac{4\pi}{3ZR^\lambda} \sqrt\frac{B(E\lambda)\uparrow}{e^2},
\end{equation} 
where $R = 1.2A^{1/3}fm$, and ${B(E\lambda)\uparrow}$ is in units of $e^2 b^2$. For $\lambda$ = 2, the values of ${B(E2)\uparrow}$ \cite{rama01} are experimental and independent of nuclear models. On the other hand, the parameter $\beta_\lambda$ depends upon the nuclear model and determines the calculation of deformation parameter. 

In the present study, we aim to understand the effect of each rotational energy level up to 12$^+$ levels i.e., 2$^+$, 4$^+$, 6$^+$, 8$^+$, 10$^+$ and 12$^+$ in the enhancement of the fusion cross-sections at below barrier energies. The calculations for the fusion cross-sections have been exercised in steps (increment of $2^+$ state in each step) from $0^+$ to $12^+$ channels.  Nonetheless, we observed that for negative and positive $\beta_4$ values, beyond 6$^+$ and 10$^+$ respectively, the higher-order channels cease to contribute significantly towards the fusion cross-section around and below the Coulomb barrier. Further, two different conditions based on the shape of nuclei: (1) $\beta_4 < 0$, and (2) $\beta_4 > 0$ are considered. The quadrupole ($\beta_2$) and/or hexadecapole ($\beta_4$) deformations of deformed nuclei are generally taken into account while characterizing the rotation of the nuclei. It has been suggested that the reactions containing nuclei with $\beta_4$ show an enhancement in the fusion cross-section at energies below the barrier \cite{fer1991,leig88}. Our present calculations assume the $^{16}$O (projectile) as spherical   \cite{fer1991,Gaut19,rajb16,rajb14,Sorl08,Greg86} to determine the rotational effect of the target nucleus on fusion cross-section effectively. It is important to note here that the nucleon transfer channels are not considered in the present calculations, which may play a significant role at energies below the Coulomb barrier \cite{deni00,lagy97,trip01}.

\subsection{For Hexadecapole deformation \texorpdfstring{$\beta_4 < 0$}{Lg}}
\label{A} \noindent
The near barrier and sub-barrier fusion cross-sections for $^{16}$O+$^{182,184,186}$W, $^{176,180}${Hf}, $^{174,176}${Yb} systems have been analyzed systematically by employing CCFULL code using the static Woods-Saxon potential with $\beta_2 > 0$, $\beta_4 < 0$ of the target nuclei. The Woods-Saxon parameterizations of Aky$\ddot{u}$z-Winther potential (AW), the values of the deformation parameters ($\beta_2,\ \beta_4$) and the excitation energy corresponding to the first excitation state are given in Table \ref{table1}. The values of the Woods-Saxon potential parameters ($V_0, r_0$ $\&$ $a_0$) are chosen to fit the experimental fusion cross-section at the above barrier energies for the case of the 1-D barrier penetration model. The variation of the total interaction potential at $\ell=0 \hbar$ with the separation distance `r' for these systems is also shown in Fig. \ref{fig. 1}. Here the total interaction potential is the sum of Woods-Saxon potential in Eq. (\ref{ws}), and the Coulomb potential $V_C$=$\frac{Z_P Z_T e^2}{r}$. From the figure, one can notice that the pocket formed for $^{16}${O}+$^{180}${Hf} reaction is much deeper in comparison to the others, which may result in a larger fusion probability. \\
\begin{figure*}
\begin{center}
\includegraphics[width=165mm,height=185mm,scale=1.5]{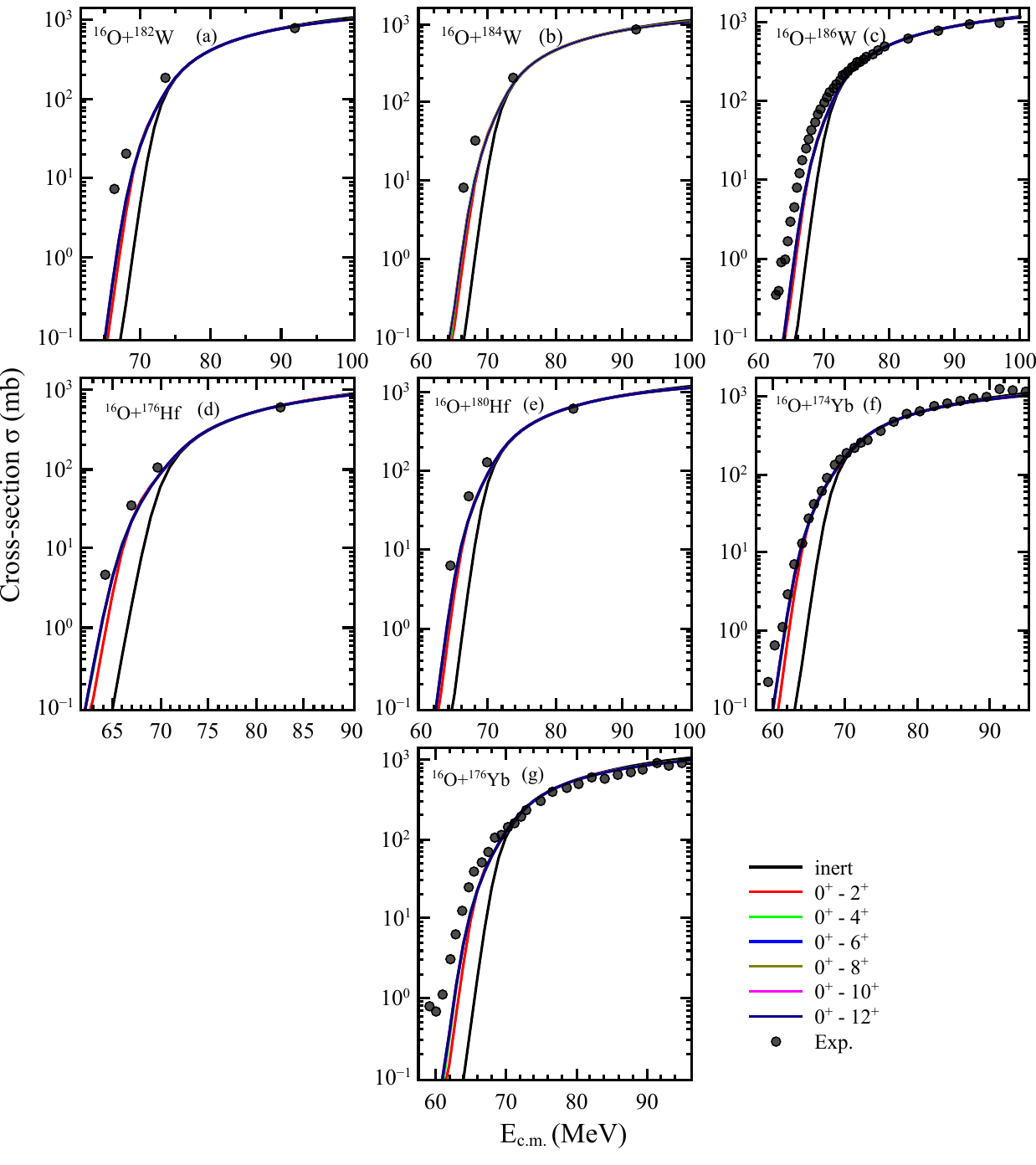}
\vspace{-0.3cm}
\caption{\label{fig2} (Color online) The fusion cross-sections estimated up to 12$^+$ channels in sequential manner as a function of $E_{c.m.}$(MeV) for $^{16}$O+$^{182,184,186}$W, $^{16}$O+$^{176,180}$Hf, and $^{16}$O+$^{174,176}$Yb nuclear reactions having quadrupole deformation and compared with experimental data \cite{rajb16,leig88,trot05}. See the text for details.}
\end{center}
\end{figure*}
The fusion cross-sections have been calculated using a 1D barrier penetration model for these reactions. The one-dimensional barrier penetration data is represented by a solid black line as shown in Fig. \ref{fig2}. From the figure, we found the obtained results underestimate the experimental data, especially at low barrier energies. In order to address the fusion cross-section of the above-mentioned reactions, rotational degrees of freedom have been included in the CCFULL calculations. Only the quadrupole deformation ($\beta_2$) is initially considered, and the calculations are performed along with various values of $\beta_2$ ranging from 0.221 to 0.324. The deformation parameters used in these reactions decreases as the mass number increases. Up to $12^+$ channels are incorporated in each system to observe the effect of each rotational level in the enhancement of the fusion cross-section and/or to predict the number of channels that are good enough to converge to the experimental data and beyond it the higher order ceases to contribute. Within the addition of rotational channels corresponding to these levels, the enhancement in the fusion cross-sections has been observed as compared to the single barrier penetration case. The contribution of the rotational energy levels on the fusion cross-section up to $6^+$ state has been found to be significant. It has been observed that the ground state quadrupole deformation alone is unable to reproduce the experimental data across the below barrier energy region. \\
\begin{figure*}
\begin{center}
\includegraphics[width=165mm,height=185mm,scale=1.5]{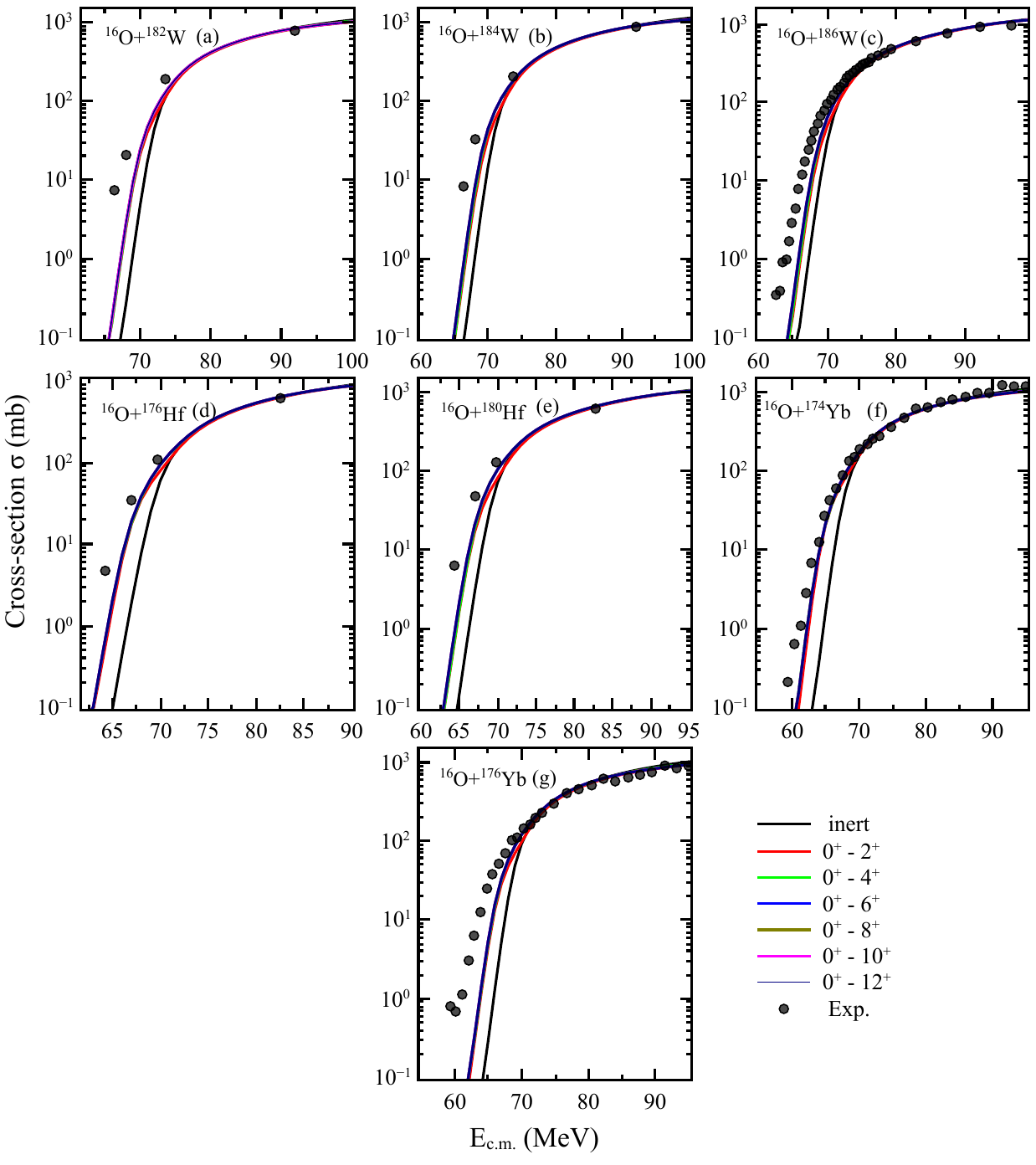}
\vspace{-0.3cm}
\caption{\label{fig3} (Color online) Same as Fig. \ref{fig2}, but with the inclusion of -ve hexadecupole ($\beta_4$) deformation}
\end{center}
\end{figure*}
To address these issues, the hexadecapole deformation ($\beta_4$) is included in the CC calculations, as illustrated in Fig \ref{fig3}. The estimated results using 1-D BPM are comparatively lower than the experimental data. The coupled channel computations are performed for this system including rotational energy levels up to the $12^+$ state of the target nuclei. A significant change in the fusion cross-section is observed for $0^+$ to $2^+$ state as compared to the 1-D BPM, and a small change for $2^+$ to $6^+$ states. Furthermore, we observe that the cross-section with the inclusion of higher-order channels, beyond $6^+$, overlaps with that of up to the $6^+$ state. The fusion cross-section for $^{16}$O + $^{182,184,186}$W reactions are calculated with the $\beta_4$ values ranging from -0.066 to -0.095. The calculated results are shown in Fig. \ref{fig3}(a), \ref{fig3}(b), \ref{fig3}(c), and the experimental data \cite{leig88,trot05} are given for comparison. The theoretical results of  CC calculations with $6^+$ channels enhance the fusion cross-section more in comparison to the 1-D BPM for $^{16}$O + $^{182}$W reaction, as represented in the Fig. \ref{fig3}(a). There is substantial variation in the $\beta_4$ value as the mass number increases from $^{182}$W to $^{184}$W i.e. the difference between their $\beta_4$ values is 0.027. As a result, there is slight change in the cross-section as we move from $2^+$ to $6^+$ channels at -0.093 and -0.095 values of the $\beta_4$ parameter, as illustrated in Fig. \ref{fig3}(b) and \ref{fig3}(c). For the above barrier energies, the calculated cross-section is a reasonably good match with the experimental data \cite{leig88,trot05}.  \\  
Similar observation can be pointed out for the case of $^{16}$O+$^{176}$Hf and $^{16}$O+$^{180}$Hf, where the $\beta_4$ values are -0.057 and -0.068, respectively. As demonstrated in Fig. \ref{fig3}(d) for $^{16}$O+$^{176}$Hf reaction, energy levels up to $4^+$ channels (solid green line) show contribution in the increment of the cross-section, whereas up to $6^+$ levels (solid blue line) are good enough for the enhancement of the fusion cross-sections for $^{16}$O+$^{180}$Hf in comparison to the 1-D BPM as shown in Fig. \ref{fig3}(e). The results obtained for $^{16}$O+$^{176}$Hf and $^{16}$O+$^{180}$Hf reactions provide a satisfactory fit to the data well above barrier experimental values \cite{leig88}. As shown in Fig. \ref{fig3} (f), rotational energy levels up to $4^+$ (solid green line) contribute to enhancing the fusion cross-section for the $^{16}$O+$^{174}$Yb reaction. In contrast, up to $6^+$ channels (solid blue line) contribute to an increase in the cross-section for the $^{16}$O+$^{176}$Yb reaction, as shown in Fig. \ref{fig3}(g). The obtained results are good enough for the convergence of available experimental data \cite{rajb16} mainly at the above barrier energies. However, at below and near-barrier energies there is no discernible change in fusion cross-section after the inclusion of rotational energy levels beyond the $6^+$ state. The above observation suggests that the fusion cross-sections are strongly influenced by the negative $\beta_4$ values as predicted in Refs. \cite{fer1991,leig88}. Up to $6^+$ state has a considerable effect on the fusion cross-section in the case of $\beta_4$ deformation; however, the negligible impact can be observed for the rest of the channels. 
\begin{figure}
\begin{center}
\includegraphics[width=75mm,height=90mm,scale=1.5]{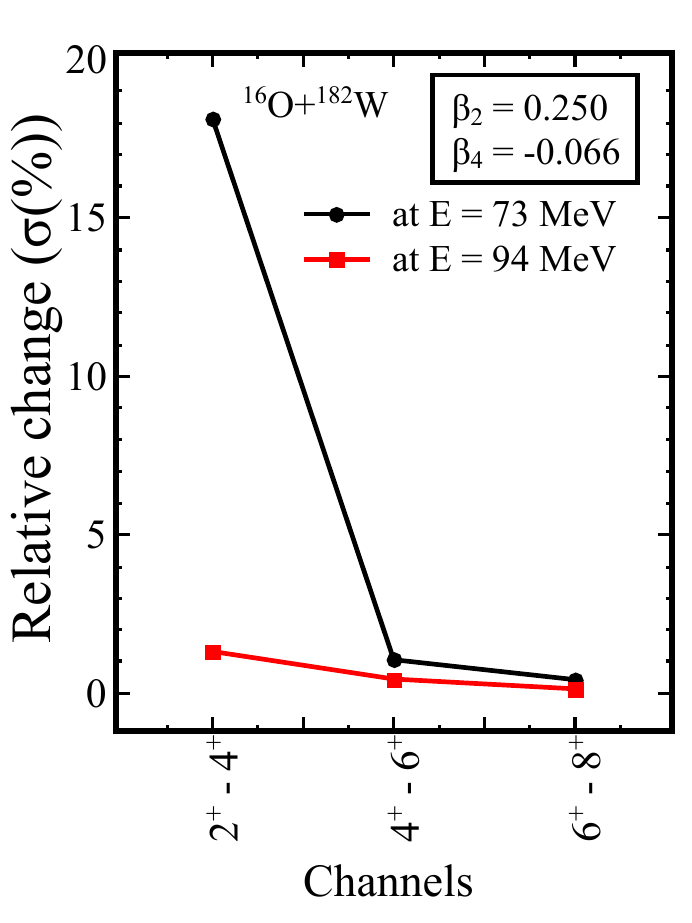}
\vspace{-0.3cm}
\caption{\label{fig4} (Color online) The relative change in the fusion cross-section w.r.t. different channels for $^{16}$O+$^{182}${W}.}
\end{center}
\end{figure}
Furthermore, for all discussed systems, the $\beta_4$ values range from -0.042 to -0.095. In addition, the result shows that if $\beta_4$ lies in the range of -0.042 to -0.057, up to $4^+$ levels show contribution in the enhancement of the cross-section w.r.t 1-D BPM. The rotational levels up to $6^+$ state gives an acceptable fit between -0.066 and -0.095 range at above barrier energies. At -0.093 and -0.095 values of the deformation parameter, there is a considerable difference in the fusion cross-section between $0^+$, $2^+$, and $4^+$ levels. Thus in general, with an increase in the magnitude of -ve $\beta_4$, there is an addition of a level or two, which starts contributing to the fusion cross-section.The relative change in the fusion cross-section w.r.t. rotational channels has been plotted in Fig. \ref{fig4} to thoroughly investigate the effect of individual channels on the cross-section corresponding to $^{16}$O+$^{182}$W reaction. As the case illustrated for $^{16}$O+$^{182}$W reaction, the relative change in the fusion cross-section of 1.06 $\%$ and 0.45 $\%$ has been observed at E = 73 MeV, 94 MeV corresponding to $4^+$ - $6^+$ state. On the other hand, the relative change in the cross-section observed is less than 1$\%$ at different $E_{c.m.}$ corresponding to $6^+$ - $8^+$ state. It shows that the rotational channel has a considerable impact on the fusion cross-section up to the $6^+$ state, whereas the other higher channels have a negligible effect. Similarly, the decreasing trend is noticed in the rest of the reactions (not shown here). \\

\subsection{For Hexadecapole deformation \texorpdfstring{$\beta_4 > 0$}{Lg}}
\begin{figure}
\begin{center}
\includegraphics[width=80mm,height=90mm,scale=1.5]{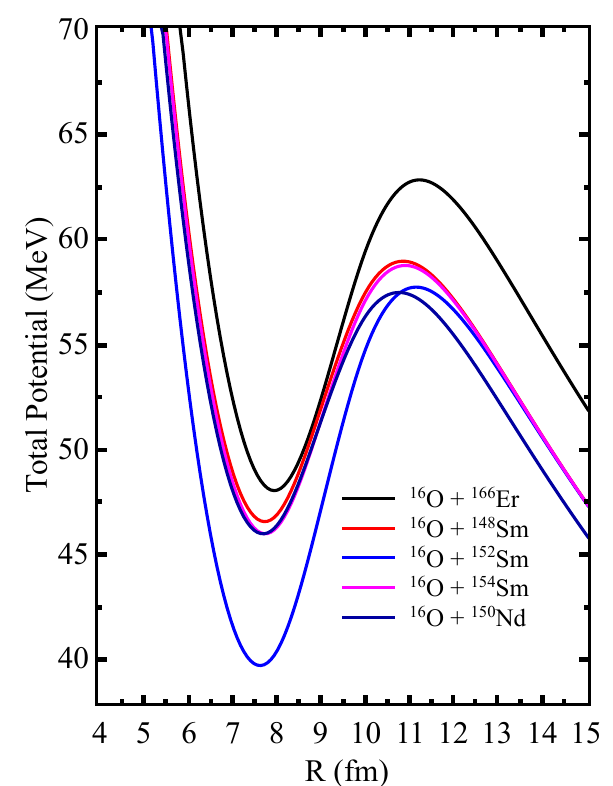}
\vspace{-0.3cm}
\caption{\label{fig5} (Color online) Same as Fig. \ref{fig. 1} but for $^{16}$O+$^{166}$Er, $^{16}$O+$^{148,152,154}$Sm, and $^{16}$O+$^{150}$Nd reactions.}
\end{center}
\end{figure}

The same procedure as discussed in the previous section \ref{A} is followed to calculate the fusion cross-sections for $^{16}$O+$^{166}${Er}, $^{16}$O+$^{148,152,154}${Sm}, $^{16}$O+$^{150}${Nd} reactions having $\beta_2 > 0$, $\beta_4 > 0$ values using the static Woods-Saxon potential by employing the CCFULL code. The Woods-Saxon parameterizations of Aky$\ddot{u}$z-Winther potential (AW), deformation parameters $\beta_2$, $\beta_4$ and the excitation energy corresponding to the first excitation state are mentioned in Table \ref{tab2}. The values of AW potential parameters are chosen to fit the experimental data at the above barrier energies for the inert case or 1-D BPM. The variation of the total interaction potential i.e. the sum of the Woods-Saxon and Coulomb potentials, at $\ell=0\hbar$ with the separation distance `r' for these systems is shown in Fig. \ref{fig5}.  In comparison with the other reactions, the pocket formed in $^{16}$O+$^{152}${Sm} reaction is substantially deeper. The probability of fusion is expected to be very significant for such relatively deeper pockets.

\begin{figure*}
\begin{center}
\includegraphics[width=165mm,height=125mm,scale=1.5]{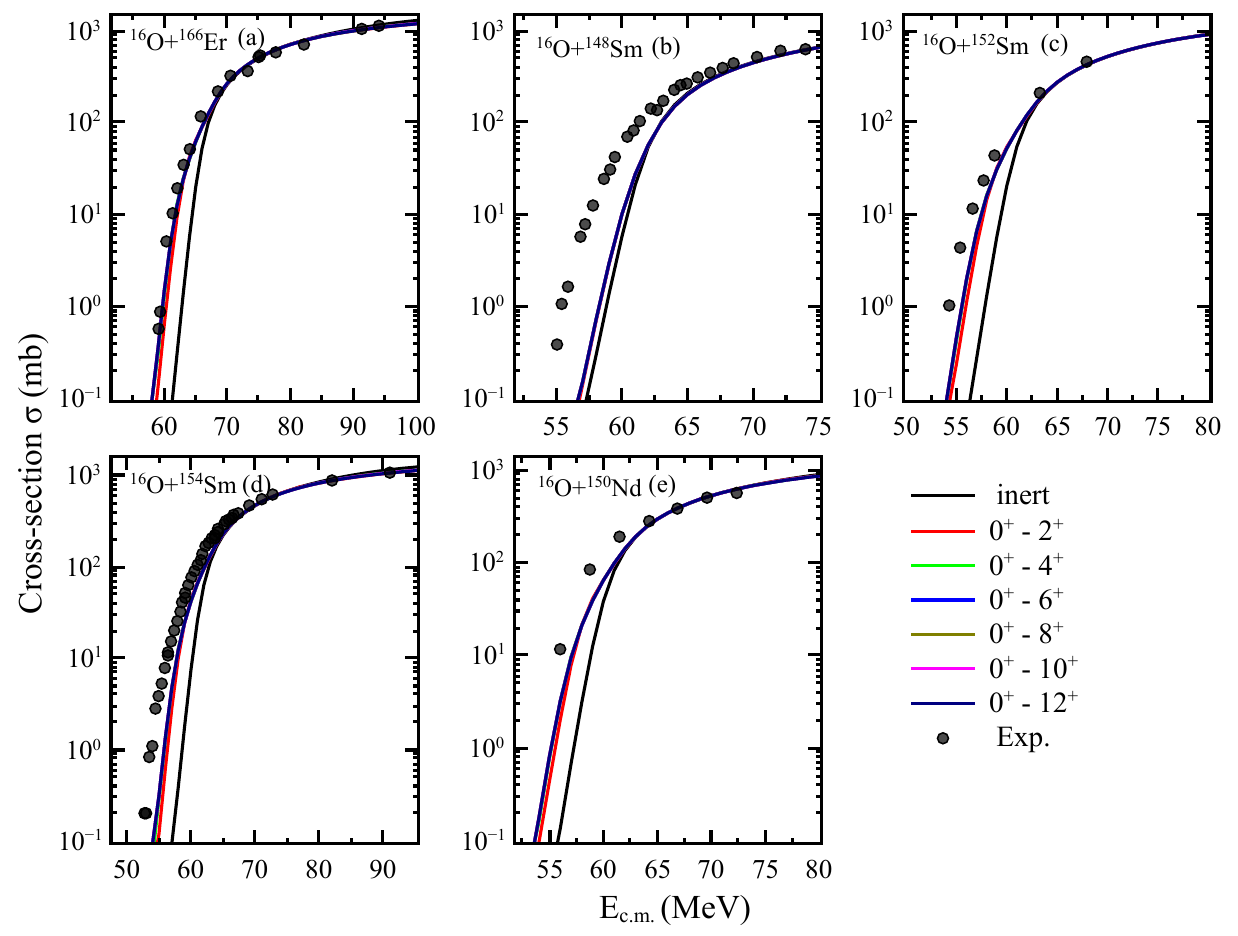}
\vspace{-0.3cm}
\caption{\label{fig6} (Color online) Same as Fig. \ref{fig2} but for $^{16}$O+$^{166}$Er, $^{16}$O+$^{148,152,154}$Sm, and $^{16}$O+$^{150}$Nd nuclear reactions and compared with experimental data \cite{stok80,wei1991,brod75}. See the text for details.}
\end{center}
\end{figure*}

The solid black line in Fig. \ref{fig6} represents the 1-D penetration case. From the figure, one can observe that at below-barrier energies, the theoretical cross-section obtained using 1-D BPM underestimates the experimental values. As mentioned earlier, rotational channels are taken into account to reduce the fusion hindrance at the below-barrier energies. Initially, the calculations are performed by considering the $\beta_2$ values ranging from 0.142 to 0.342. The contribution of the rotational energy levels in the enhancement of the fusion cross-section obtained is the same as in the previous Section (\ref{A}) except for $^{16}$O+$^{148}$Sm nuclear reactions. In these reactions, the target nuclei have $\beta_2$ value 0.1423, whereas, in other reactions, the target nuclei are highly deformed (0.285 $\leq \beta_2 \leq$ 0.342). Based upon these values, the rotational levels up to $4^+$ show enhancement in the fusion cross-section as compared to the 1-D barrier penetration model as shown in Fig. \ref{fig6}(b). However, higher-order channels have a negligible contribution toward the fusion cross-section. The ground state $\beta_2$ values alone as presented in Fig. \ref{fig6} are incapable of reproducing the experimental data over the whole energy range, thus need to be included in the $\beta_4$ along with $\beta_2$.
\begin{table} 
\caption{\label{tab2} Same as Table \ref{table1}, but for the case of ($\beta_2 > 0, \beta_4 > 0 $)}
\centering
{\begin{tabular}{ccc|ccc}
\hline \hline 
System & $V_0 (MeV)$ & $r_0 (fm)$ & \multicolumn{3}{c}{Target} \\
        &         & & $E_2^+ (MeV)$ & $\beta_2$ & $\beta_4$ \\
\hline
$^{16}$O+$^{166}$Er& 67.0& 1.185&0.080& 0.342& 0.007\\
  $^{16}$O+$^{148}$Sm& 62.204& 1.177& 0.550& 0.1423& 0.060\\
  $^{16}$O+$^{152}$Sm&70.0&1.18&0.121&0.3064& 0.097\\
  $^{16}$O+$^{154}$Sm&62.53&1.17& 0.081&0.341& 0.105\\
  $^{16}$O+$^{150}${Nd}&60.0&1.165& 0.130&0.2853& 0.110\\
  \hline
\end{tabular}}
\end{table}
\begin{figure*}
\begin{center}
\includegraphics[width=165mm,height=125mm,scale=1.5]{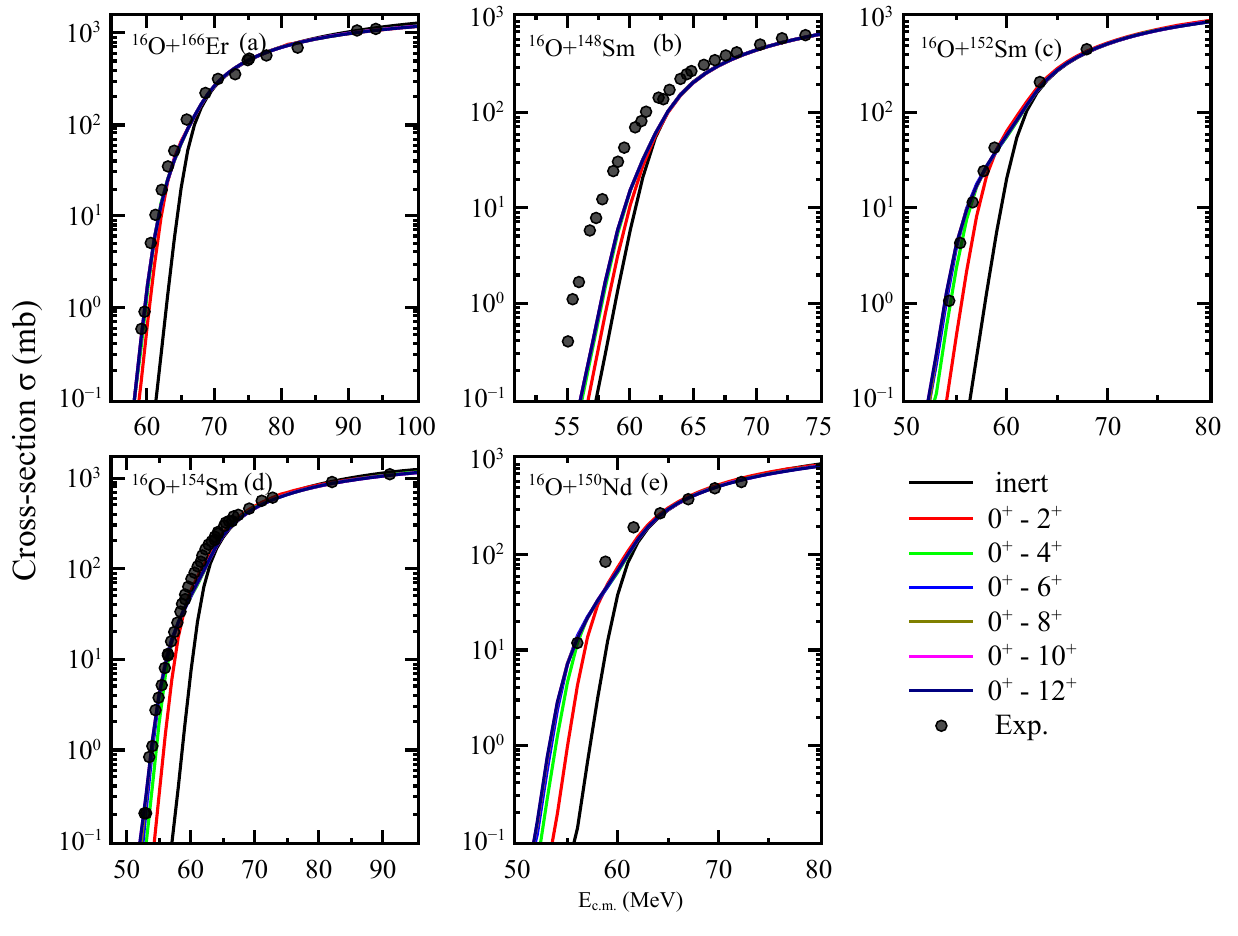}
\vspace{-0.3cm}
\caption{\label{fig7} (Color online) Same as Fig. \ref{fig6}, but with the inclusion of $-$ve hexadecupole ($\beta_4$) deformation}
\end{center}
\end{figure*}

The positive $\beta_4$ plays an important role in the enhancement of the fusion cross-sections mainly at below-barrier energies. For $^{16}$O+$^{166}${Er}, there is an enhancement in the fusion cross-sections up to $6^+$ levels (solid blue line) and other higher channels up to $12^+$ have negligible impact on fusion cross-sections to converge towards the experimental data \cite{fer1991} as shown in Fig. \ref{fig7}(a). The difference between $4^+$ and $6^+$ channel is quite significant because of strong $\beta_2$, and $\beta_4$ value for $^{166}${Er} target nuclei. For Sm targets, there is a significant variation in the quadrupole deformation $\beta_2$ ranging from 0.1423 to 0.306 and also in the value of $\beta_4$ from 0.060  to 0.097 for $^{148}${Sm} and $^{152}${Sm}. For $^{16}${O}+$^{148}$Sm reaction, the effect of hexadecapole deformation on the fusion cross-section obtained is the same. Similarly, the outcomes of $^{16}${O}+$^{152}${Sm}, and $^{16}${O}+$^{154}$Sm reactions are also identical. The $6^+$ channels (solid blue line) contribute to the enhancement of the fusion cross-section in $^{16}${O}+$^{148}${Sm} reaction, as shown in Fig. \ref{fig7} (b), and \ref{fig7} (c). In contrast, the $10^+$ channels (solid magenta line) play a significant role in the increment of the fusion cross-section or reduced fusion hindrance at below barrier energies in $^{16}${O}+$^{152}${Sm}, and $^{16}${O}+$^{154}${Sm} reactions, as shown in Fig. \ref{fig7}(c), and \ref{fig7}(d). Theoretical results obtained for these reactions give the best fit with the experimental data \cite{leig95,stok80,wei1991}. This difference in the rotational energy levels is due to $\beta_4$ values because these values change significantly from 0.060 to 0.097. The significant change between each channel are observed because of strong deformation ($\beta_2, \beta_4$) values in case of $^{154}${Sm}. Also, it is well known that $^{154}${Sm} is a perfect rotor \cite{wei1991}. Further, with the inclusion of higher channels, a negligible effect on the fusion cross-sections is observed.

The similar results are obtained for $^{16}$O+$^{150}${Nd} reactions system.  In the case of $^{16}$O+$^{150}${Nd}, $10^+$ (solid magenta line) channels play a significant role in increasing the cross-section, as illustrated in Fig. \ref{fig7}(e). These theoretical results provide a satisfactory fit with the experimental values \cite{brod75}. The involvement of higher-order channels up to $12^+$ does not affect the fusion cross-sections. There is a significant difference between different channels w.r.t. 1-D barrier penetration model because of +ve deformation values. From the above results, we conclude that the rotational levels up to $4^+$ are good enough to converge the experimental data except for the reaction in which $\beta_2$ values are less than 0.142. The value of the $\beta_4$ parameter for this system is 0.060. We can conclude from the above-discussed systems, that when the value of $\beta_4$ lie in a range between 0.007-0.060, rotational energy levels up to $6^+$ lead to an enhancement in the fusion cross-section. However, when the $\beta_4$ value lies in the range 0.097-0.110, $10^+$ levels play a significant role in increasing the cross-section and also provide a satisfactory fit with the experimental values.

\begin{figure}
\begin{center}
\includegraphics[width=75mm,height=90mm,scale=1.5]{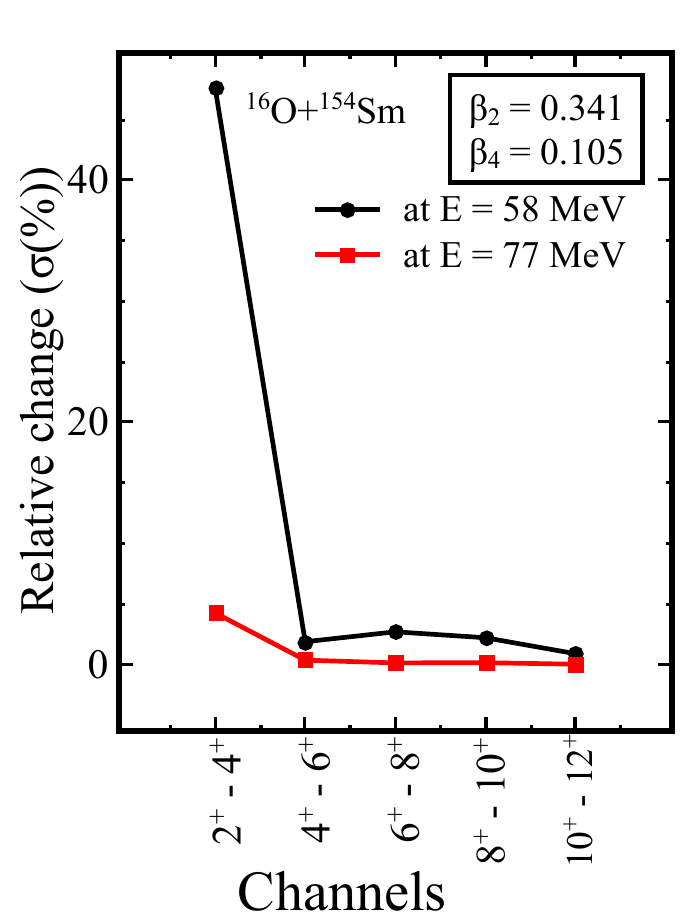}
\vspace{-0.3cm}
\caption{\label{fig8} (Color online) Same as Fig. \ref{fig4} but for $^{16}$O+$^{154}${Sm}.}
\end{center}
\end{figure}

As illustrated for $^{16}$O+$^{154}${Sm} reaction , Fig. \ref{fig8} depicts the relative change in the fusion cross-section with respect to rotational channels. The relative change in the fusion cross-section at E = 58 MeV (below the Coulomb barrier region) and 77 MeV (above the Coulomb barrier region) of energy is 2.21 $\%$ and 0.15 $\%$, respectively, corresponding to the $8^+$ - $10^+$ state. In contrast, at different $E_{c.m.}$ corresponding to $10^+$ - $12^+$ states, the relative change in the cross-section found is less than 1$\%$. However, Rowley \emph{et al.} \cite{rowl91} had demonstrated that up to 3-channels of the rotational energy levels are enough to address the experimental data for $^{154}${Sm}. Nonetheless, certain discrepancies were evident between the experimental and the theoretical results. It was suggested that with the inclusion of phonon(s) or transfer channels, these discrepancies can be overcome. Comparatively, here the relative change clearly shows that up to $10^+$ states of the rotational channel have a significant effect on the fusion cross-section. Moreover, our results indicate that when higher-order channels are used, there is no need to include the phonon and/or the transfer coupling for reproducing the experimental data.  The rest of the reactions follow the same pattern as the first in terms of decreasing intensity.

\subsection{Fitting Curve}
The development of the radioactive beam makes it possible to synthesize a variety of nuclei that lie in the valley of stability as well as far from the $\beta$-stability region including the superheavy island. Determining the proper reaction dynamics necessitates a thorough understanding of the synthesis and characteristics of these nuclei. Many efforts are being devoted to the direction of theoretical and experimental studies. On the other hand, experimental verification is too difficult. As a result, we need to execute theoretical modelling to confirm their characteristics in terms of reaction dynamics. As we know, CCFULL is one of the computational codes used to study fusion dynamics, which requires a detailed description of Woods-Saxon nuclear potential parameters, i.e., $V_0$ and $r_0$ of the colliding nuclei. It is difficult to extract the potential parameters for the recently synthesized or predicted target(s) nuclei with $^{16}O$ induced reactions. As a result, for interacting nuclei that lie in the stable and unstable mass region, one must fit the free parameters for the nuclear potential in CCFULL. Among the considered reactions, the fusion cross-sections for twelve reaction system, namely,  $^{182,184,186}$W, $^{176,180}${Hf}, $^{174,176}${Yb}, $^{166}${Er}, $^{148,152,154}${Sm}, $^{150}${Nd} are in good agreement with the available experimental data at above barrier energies.

\begin{figure}
\begin{center}
\includegraphics[width=85mm,height=110mm,scale=1.7]{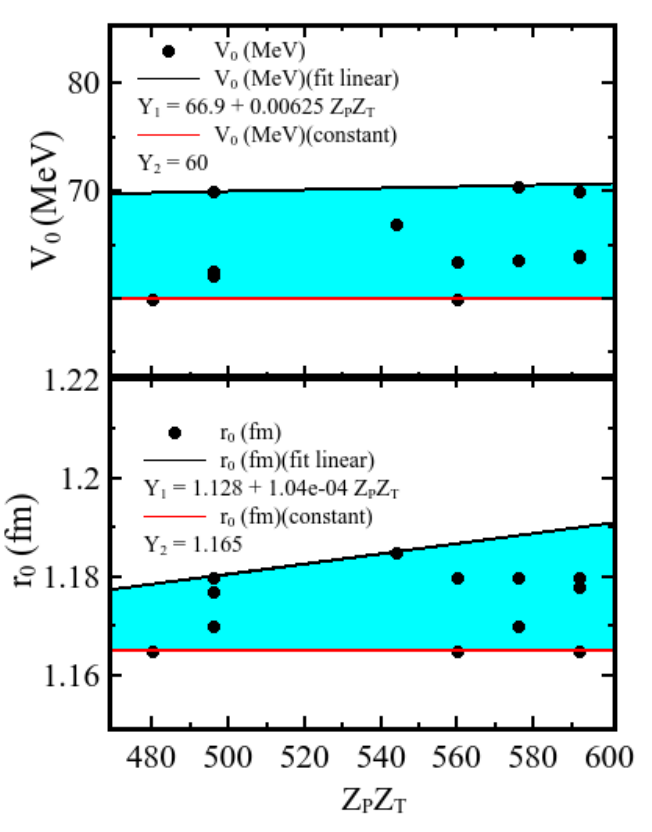}
\vspace{-0.3cm}
\caption{\label{fig9} (Color online) Variation of fitted $V_0$ and $r_0$ as a function of $Z_pZ_T$. The fitted polynomial are also shown.}
\end{center}
\end{figure}

In Woods-Saxon parameterization potential, the standard value of diffuseness parameter $a_0$ = 0.65 fm and the value of the parameters such as $V_0$ and $r_0$ are difficult to extract for the unexplored reaction systems. The Woods-Saxon parameters available on the NRV website \cite{nrv} are unable to fit the experimental data even at the above barrier energies. Thus the values of $V_0$ and $r_0$ used in this study are chosen to fit the experimental data at above barrier energies for the inert case and later on the coupling to various rotational energy levels is included. The motivation of the present study is to extract the relative contribution of individual rotational energy levels up to higher-order states (${12}^+$). Also, the results drawn are independent of the choice of nuclear potential. In this direction, we have given an algebraic function by a linear fitting of the curve for known systems as a function of $Z_PZ_T$ as shown in Fig. \ref{fig9}. Thus, one can generate the value of $V_0$ and $r_0$ using the algebraic formula for $V_{02} = 66.9 + 0.00625 Z_p Z_t$, and $V_{01}$ = 60 MeV and $r_{02} = 1.128 + 1.04*10^{-4}Z_p Z_t$, and $r_{01}$ = 1.165fm. Here the subscript {\it 1}, and {\it 2} stands for the lower and upper limit of the extracted band region for $V_0$, and $r_0$ as shown in Fig. (\ref{fig9}). These fitted values of $V_0$, and $r_0$ are valid from $Z_p Z_t$ = 480 to $Z_p Z_t$ = 592 in the rotational region of the Periodic Table. Using these simple algebraic formulas, one can extract the potential parameters for the limited range of target nuclei interacting with $^{16}O$ as a projectile, which will be proven essential for the upcoming experiment.

\section{Summary and Conclusions}
\label{result} \noindent
In the present work, we have studied the effect of individual rotational energy levels on the fusion cross-section at deep sub-barrier energies in heavy-ion nuclear reactions. Here, we have considered $^{16}O$-induced reactions in which target nuclei chosen are rotational in nature(i.e. $^{182,184,186}$W, $^{176,180}${Hf}, $^{174,176}${Yb}, $^{166}${Er}, $^{148,152,154}${Sm}, $^{150}${Nd}) and projectile is spherical. As such, we have demonstrated the effect of nuclear shapes on fusion cross-sections by considering the deformed target nuclei (rotational) only. For different values of deformation parameters, the role of different rotational energy levels has been described in the terms of the fusion cross-sections. The contribution of the rotational energy levels up to $6^+$ levels has been observed on the fusion cross-section for quadrupole deformation ($\beta_2$). For -ve hexadecapole deformation, higher-order channels up to $6^+$ are found suitable for the convergence of the cross-sections towards experimental data whereas for +ve $\beta_4$ deformation, $10^+$ levels fit the data well. It is noticed that channels beyond $10^+$ have a negligible impact on the fusion cross-section. For the determination of the free parameter of Woods-Saxon potential, the parameters are fitted as an algebraic function of $Z_PZ_T$. This work will be provided in identifying the combinations of the target nuclei with $^{16}O$ projectile within the range of $Z_PZ_T$ from 480 to 592. 
\section*{Acknowledgements}
This work has been supported by Science Engineering Research Board (SERB), File No. CRG/2021/001229, FOSTECT Project No. FOSTECT.2019B.04, FAPESP Project No. 2017/05660-0. 


\end{document}